\documentclass[aps,pra,10pt,superscriptaddress,onecolumn,nofootinbib]{revtex4-2}
\usepackage{graphicx}
\usepackage{amsmath,amsthm,amssymb,dsfont}
\usepackage{hyperref}
\usepackage{mdframed}
\usepackage{orcidlink}
\usepackage{soul}
\newcommand{\ket}[1]{\left| #1 \right\rangle}
\newcommand{\bra}[1]{\left\langle #1 \right|}
\newcommand{\ketbra}[2]{\left|#1\right\rangle\hskip-1mm\left\langle #2\right|}
\newcommand{\braket}[2]{\langle #1|#2 \rangle}
\soulregister\ref{7}
\soulregister\cite{7}

\begin{document}

\title{Is the dynamical quantum Cheshire cat detectable?}

\author{Jonte R. Hance\,\orcidlink{0000-0001-8587-7618}}
\email{jonte.hance@newcastle.ac.uk}
\affiliation{School of Computing, Newcastle University, 1 Science Square, Newcastle upon Tyne, NE4 5TG, UK}
\affiliation{Quantum Engineering Technology Laboratories, Department of Electrical and Electronic Engineering, University of Bristol, Woodland Road, Bristol, BS8 1US, UK}
\author{James Ladyman}
\affiliation{Department of Philosophy, University of Bristol, Cotham House, Bristol, BS6 6JL, UK}
\author{John Rarity}
\affiliation{Quantum Engineering Technology Laboratories, Department of Electrical and Electronic Engineering, University of Bristol, Woodland Road, Bristol, BS8 1US, UK}

\begin{abstract}
    We explore how one might detect the dynamical quantum Cheshire cat proposed by Aharonov et al. We show that, in practice, we need to bias the initial state by adding/subtracting a small probability amplitude (`field') of the orthogonal state, which travels with the disembodied property, to make the effect detectable (i.e. if our initial state is $\ket{\uparrow_z}$, we need to bias this with some small amount $\delta$ of state $\ket{\downarrow_z}$). This biasing, which can be done either directly or via weakly entangling the state with a pointer, effectively provides a phase reference with which we can measure the evolution of the state. The outcome can then be measured as a small probability difference in detections in a mutually unbiased basis, proportional to this biasing $\delta$. We show this is different from counterfactual communication, which provably does not require any probe field to travel between sender Bob and receiver Alice for communication. We further suggest an optical polarisation experiment where these phenomena might be demonstrated in a laboratory.
\end{abstract}

\maketitle

\section{Introduction}
In 2013, Aharonov et al first described the quantum Cheshire cat phenomenon. This is where they claim a property of a particle (e.g. the weak value of its polarisation\footnote{There is a lot of controversy around weak values, but we will not address this in this paper.}) is disembodied from that particle, and shown to travel through regions the particle could not traverse \cite{Aharonov2013Cheshire}. It was named in reference to the Cheshire Cat in Alice in Wonderland, who could slowly fade away, leaving only his disembodied grin in his place. {The protocol was later demonstrated experimentally \mbox{\cite{Denkmayr2014QCCExp,Ashby2016QCCExperiment,sponar2016fundamental,sponar2018weak,Kim2021,nawaz2019atomic,Danner2023Neutrons,Sahoo2023}}, and built upon, with recent work claiming to allow the swapping of this disembodied property between two particles \mbox{\cite{Das2020QCCSwitch,Liu2020QCCExchangeExp,Danner2023Neutrons}}, the delayed choice of which path carries the particle and which carries the disembodied property \mbox{\cite{Das2021Delayed,Wagner2023Delayed}}, the disembodying of multiple properties simultaneously \mbox{\cite{pan2020multi,danner2023three}}, the splitting of the mass from the momentum of a quantum particle \mbox{\cite{waegell2024separating}}, and even the separation of the ``wave-particle duality'' of a particle \mbox{\cite{li2023experimental}}.}

More recently, Aharonov et al gave a protocol where they claim this disembodied property (in their example, the weak value of spin) can be altered while separated from its possessing particle - a dynamical Cheshire cat \cite{Aharonov2021Dynamical}. They further say that this dynamical quantum Cheshire cat could be behind the puzzling phenomenon of counterfactual communication \cite{Salih2013Protocol,Salih2018Laws,Aharonov2019Modification,Aharonov2020Nonlocal,Hance2021Quantum} (and related forms of counterfactual (quantum) information transfer/metrology \cite{Salih2016Qubit,Salih2018Paradox,Salih2020DetTele,Salih2021EFQubit,Hance2021CFGI}). In this paper, we show how this phenomenon can be detected experimentally. We note however that this involves adding a small orthogonal probe field, which travels with the property, which can measure the evolution of the state. The measured weak value remains constant as we reduce the value of this orthogonal probe field, although its uncertainty increases. Of course this value is indeterminate when the probe field is zero. The effect cannot be observed when the probe field is zero. {We therefore show that, contrary to Aharonov et al's claims, the dynamical Cheshire cat effect is inherently different from} counterfactual communication, where the transmitted information is provably not associated with a probe field passing between communicating partners.

\section{Dynamical quantum Cheshire cat}
\begin{figure*}
    \centering
    \includegraphics[width=0.75\linewidth]{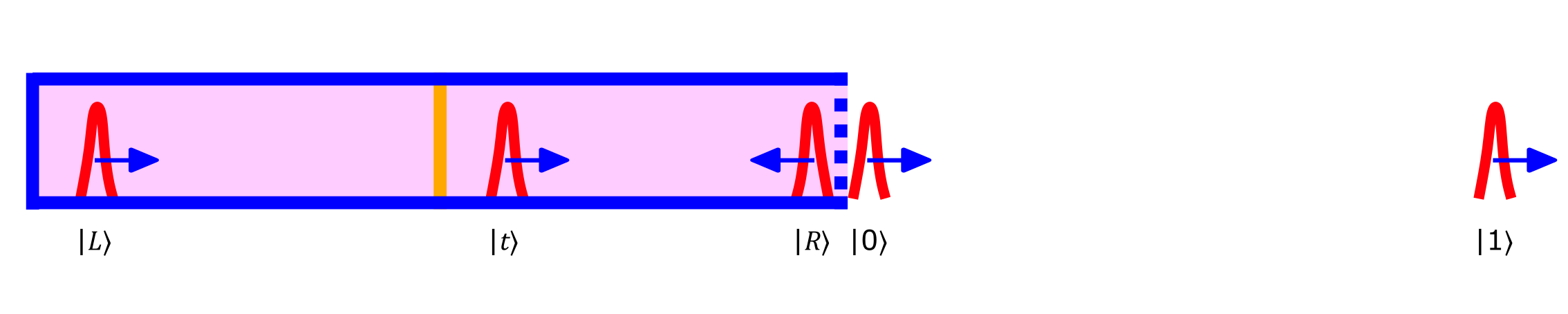}
    \caption{Aharonov et al's protocol for applying the dynamical quantum Cheshire cat. The yellow line is a semi-transparent partition with reflection-transmission ratio $1:\epsilon$, and the dotted line is a spin-dependent mirror which transmits $\ket{\uparrow_z}$ and reflects $\ket{\downarrow_z}$.}
    \label{fig:DynamCCDiagl}
\end{figure*}

In their paper \cite{Aharonov2021Dynamical}, Aharonov et al propose a protocol consisting of two coupled cavities of equal length, connected via a semi-transparent partition with transmissivity parameter $\epsilon=\pi/2N$ for large integer $N$ (see Fig. \ref{fig:DynamCCDiagl}). A single-particle wave-packet is introduced into the left-hand cavity before its left-side wall is closed, and evolves to have small amounts of field cross the semi-transparent partition and probe the spin-dependent wall closing the right-hand side of the right-hand cavity. This wall transmits field with spin $\ket{\uparrow_z}$ and reflects field with $\ket{\downarrow_z}$ spin.

The overall length of the two coupled cavities is distance $D$, which would take the wave-packet time $T$ to traverse were the cavities perfectly coupled (i.e. the particle has constant velocity $D/T$). As the two cavities are of equal length, it takes the wave-packet reflected by the semi-transparent partition time $T$ to travel from the left-hand side of the left cavity to the partition, get reflected, then return back to the left hand side of the left cavity.

The key times for us to consider are $t_n=nT$, for integer $n$. At each of these moments, there will be a wave-packet $\ket{L}$ at the leftmost wall, a wave-packet $\ket{R}$ just reflected at the spin-dependent right wall, a wave-packet $\ket{0}$ just transmitted at the spin-dependent right wall, and states $\ket{k}$ for integer $k$, $0\leq k\leq n-1$, each distance $kD$ from the spin-dependant right-hand wall.

Given we input the particle in spin state $\ket{\uparrow_z}$, the evolution between time $t_n$ and $t_{n+1}$ is given by unitary operator $\hat{U}$ where
\begin{equation}
    \begin{split}
        \hat{U}\ket{L}\ket{\uparrow_z} &= \cos{\epsilon}\ket{L}\ket{\uparrow_z} + \sin{\epsilon}\ket{0}\ket{\uparrow_z},\\
         \hat{U}\ket{k}\ket{\uparrow_z} &= \ket{(k+1)}\ket{\uparrow_z}
    \end{split}.
\end{equation}

Therefore, given the state starts at $\ket{\Psi(n=0)} = \ket{L}\ket{\uparrow_z}$, the overall state at $t=nT$ is
\begin{equation}
\begin{split}
     \ket{\Psi(n)}=\cos^n{\epsilon}\ket{L}\ket{\uparrow_z} + \sum_{k=0}^{n-1} \sin{\epsilon}\cos^k{\epsilon}\ket{(n-1-k)}\ket{\uparrow_z}.
\end{split}
\end{equation}

We now want to see what happens after large integer $n=N$. Due to the interrupted evolution from leaks through the spin-dependant wall, it would take the wave-packet time of order $\mathcal{O}(N^2T)$ to fully leak out of the box. Therefore, for big $N$, the wave-packet is still effectively in state $\ket{L}\ket{\uparrow_z}$, with infinitesimal chance of leaving (order $\mathcal{O}(1/N)$).

Ergo, the state at time $2NT$ is
\begin{equation}
\begin{split}
\ket{\Psi(2NT)} &= \hat{U}^{2N}\ket{L}\ket{\uparrow_z}\\
    &= \cos^{2N}{(\pi/2N)}\ket{L}\ket{\uparrow_z}\\
    &\approx \ket{L}\ket{\uparrow_z}.
\end{split}
\end{equation}

So far, so simple.
However, something weird happens when we consider this $\ket{\uparrow_z}$ spin decomposed as $(\ket{\uparrow_x}+\ket{\downarrow_x})/\sqrt{2}$. To see this, consider the effect of the same system on $\ket{L}\ket{\downarrow_z}$. In this case, if it crosses from $\ket{L}$ to $\ket{R}$, the wave-packet is reflected by the spin-dependent right-hand wall, and so sent back to the semi-transparent partition, where again it is affected by the transmissivity parameter $\epsilon$. However, this time, the input direction of the wave-packet must be considered to ensure the partition is unitary, requiring a phase of $-1$ be applied to anything crossing from $\ket{R}$ to $\ket{L}$.

Therefore, the evolution of $\ket{\downarrow_z}$ over time $nT$ is
\begin{equation}
    \begin{split}
        &\hat{U}^{n}\ket{L}\ket{\downarrow_z} = \cos{(n\epsilon)}\ket{L}\ket{\downarrow_z} + \sin{(n\epsilon)}\ket{R}\ket{\downarrow_z},\\
        &\hat{U}^{n}\ket{R}\ket{\downarrow_z} = \cos{(n\epsilon)}\ket{R}\ket{\downarrow_z} - \sin{(n\epsilon)}\ket{L}\ket{\downarrow_z},
    \end{split}
\end{equation}
an so, after time $2NT$, we get the state
\begin{equation}
    \hat{U}^{2N}\ket{L}\ket{\downarrow_z} = -\ket{L}\ket{\downarrow_z}.
\end{equation}

This means, if we start with state $\ket{\uparrow_x}$, after time $2NT$ we get state
\begin{equation}
\begin{split}
       \hat{U}^{2N}\ket{L}\ket{\uparrow_x} &=    \hat{U}^{2N}\ket{L}\frac{\ket{\uparrow_z} + \ket{\downarrow_z}}{\sqrt{2}}\\
       &= \ket{L}\frac{\cos^{2N}{(\pi/2N)}\ket{\uparrow_z} + \ket{\downarrow_z}}{\sqrt{2}}+\text{Loss}\\
       &\approx \ket{L}\frac{\ket{\uparrow_z} - \ket{\downarrow_z}}{\sqrt{2}}
       =\ket{L}\ket{\downarrow_x}.
\end{split}
\end{equation}

Similar evolution occurs when, starting with $\ket{\downarrow_x}$ we get out $\ket{\uparrow_x}$ after time $2NT$. This is equivalent to the (approximate) Heisenberg evolution
\begin{equation}
    \hat{U}^{\dagger,2N}\ketbra{L}{L}\sigma_x\hat{U}^{2N}\approx\ketbra{L}{L}(-\sigma_x),
\end{equation}
where $\sigma_x$ is the Pauli-$x$ operator (for spin, $\ketbra{\uparrow_x}{\uparrow_x}-\ketbra{\downarrow_x}{\downarrow_x}$) This is different to if the right-hand-cavity's right-hand-wall was just open, as that would cause no flipping of $\sigma_x$.

This leads us to ask whether we can observe this evolution when no element of the wavepacket ever returns from the right-hand side (i.e. when the initial state is $\ket{\uparrow_z}$)?

\section{Detecting the effect directly}

\begin{figure}
    \centering
    \includegraphics[width=0.75\linewidth]{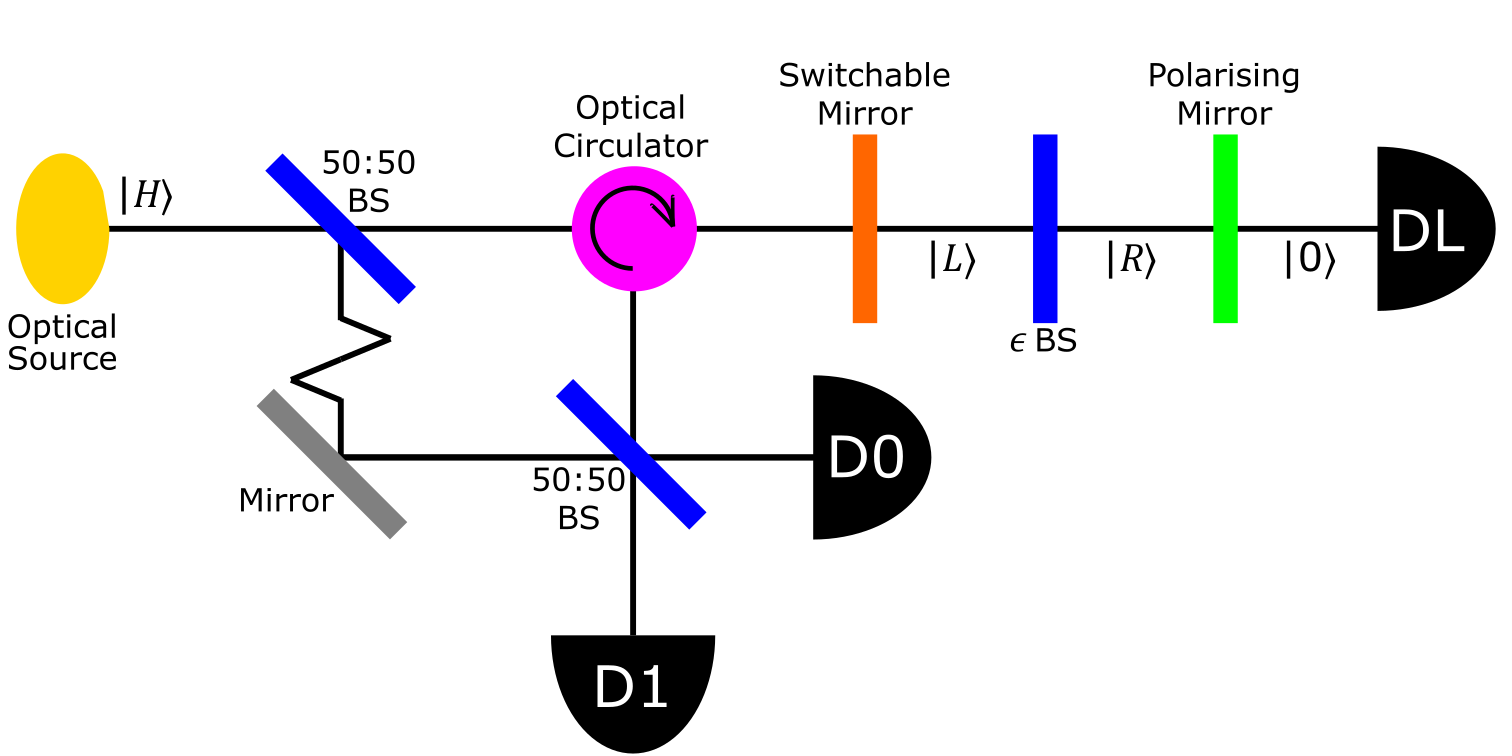}
    \caption{Aharonov et al's dynamical quantum Cheshire cat protocol, adapted to polarisation rather than spin, and with an additional homodyne detection element added. $\ket{L}$ and $\ket{R}$ indicate the left and right cavity modes, as corresponding to those in Fig. \ref{fig:DynamCCDiagl}. The switchable mirror admits the wavepacket into the protocol device (right-hand side), and keeps it there for time $2NT$ before releasing it back left for measurement. The paired balanced beamsplitters act to split a reference wavepacket from the wavepacket admitted into the protocol, before recombining the wavepackets afterwards to perform a homodyne measurement over them. The paths in the Mach-Zehnder interferometer formed by the beamsplitters are balanced so that, if there is no phase accrued by the wavepacket in the protocol, the combined wavepackets deterministically go to $D0$. Therefore, any light going to $D1$ shows the protocol creates a measurable effect.}
    \label{fig:HomodyneDynamCC}
\end{figure}

To observe this effect, we transfer this particle experiment to a simple experiment realisable in a optics laboratory. 

To do this, we first move from spin to polarisation as our degree of freedom, replacing $\ket{\uparrow_z}$ and $\ket{\downarrow_z}$ with $\ket{H}$ (horizontal) and $\ket{V}$ (vertical) polarisation respectively, and so $\ket{\uparrow_x}$ and $\ket{\downarrow_x}$ with $\ket{D}$ (diagonal) and $\ket{A}$ (anti-diagonal) polarisation. Therefore, our $\sigma_x$ operator is $\ketbra{D}{D}-\ketbra{A}{A}$. Similarly, we replace the spin-dependent mirror with a polarising mirror, which transmits $H$-polarised and reflects $V$-polarised light, and  replace the leaky mirror with a beamsplitter with reflection:transmission ratio $\cos(\epsilon):\sin(\epsilon)$.

The question we now ask is whether we can observe the $\sigma_x=-1$-evolution of the wavepacket when it is initially $\ket{H}$-polarised (and thus passes directly through the polarising mirror in cavity $R$)?

In Fig. \ref{fig:HomodyneDynamCC}, we give an apparatus able to perform Aharonov et al's protocol optically (everything to the right of the switchable mirror). We add however a way of obtaining a reference beam using a balanced beamsplitter, keeping it in a delay line while the protocol occurs, then using it to perform a homodyne measurement on the wavepacket after the protocol. This allows us to see any phase change in the wavepacket polarisation due to the protocol.

Given the protocol takes
\begin{equation}
    \ket{H}\ket{L}\rightarrow\cos^{2N}{(\pi/2N)}\ket{H}\ket{L}+\;\text{Loss}\; \approx\ket{H}\ket{L},
\end{equation}
it obviously causes no phase difference to be accrued (aside from that caused by path length difference, which can be mitigated against), and so the phase between the protocol and reference wavepackets is nil. This is the identical result to when there is no polarising mirror present in cavity $R$ (and so all light in cavity $R$ leaks out to the right, and detector $DL$). This means, as expected, the combination of the two wavepackets still always goes to $D0$, meaning the effect cannot be observed directly. This leads us to ask whether it can be observed in some indirect way.

\section{Detecting via weak value of \texorpdfstring{$\sigma_x$}{sigma-x}}

Given we just showed that the dynamical quantum Cheshire cat effect cannot be observed directly using homodyne measurement, we now see if it is observable theoretically using weak values. Note that, while calculating weak values theoretically doesn't cause the same coupling errors as weak measurement, actually obtaining a weak value experimentally causes disturbances, due to the initial weak measurement required to entangle the system with the pointer \cite{Kastner2017DemystifWeakVals,Ipsen2022Disturbance}. This can be seen in Vaidman's nested interferometer \cite{Vaidman2013Past,Hance2021WeakTrace,Vaidman2023CommWeakVals,Hance2023ReplyCommWeakVals}. Therefore, it is debatable whether this measurement actually observes the effect on an unperturbed system, or just the effect applied to some perturbation from our system of interest.

In their paper, Aharonov et al proposed demonstrating the dynamical quantum Cheshire cat effect 
by weakly measuring $\sigma_x$ at the start of the experiment, then strongly measuring it after time $2NT$, to get the weak value of $\sigma_x$ as
\begin{equation}
    \sigma_x^w\approx-1.
\end{equation}

They obtain this from the weak value formula
\begin{equation}
    \hat{A}^w = \frac{\bra{\psi_f}\hat{A}\ket{\psi_i}}{\braket{\psi_f}{\psi_i}},
\end{equation}
where in this case, $\hat{A} = \sigma_x$, $\ket{\psi_i} = \ket{H}\ket{L}$, and
\begin{equation}
\begin{split}
        \ket{\psi_f}&= \hat{U}^{2N}\ket{D}\ket{L}\\
        &= \frac{\cos^{2N}(\pi/2N)\ket{H}-\ket{V}}{\sqrt{2}}\ket{L} +\text{ Loss}\\
        &\approx\ket{A}\ket{L}.
\end{split}
\end{equation}

Therefore,
\begin{equation}\label{Eq.DQCCWeak}
\begin{split}
        \sigma_x^w &= \frac{\bra{L}\bra{A}\Big(\ketbra{D}{D}-\ketbra{A}{A}\Big)\ket{H}\ket{L}}{\bra{L}\braket{A}{H}\ket{L}}\\
        &= \frac{-1/\sqrt{2}}{1/\sqrt{2}} = -1.
\end{split}
\end{equation}

Looking more generally, we see that if we give the postselected final state as 
\begin{equation}
    \ket{\psi_f^\beta} = \Big(\cos(\beta)\cos^{2N}(\pi/2N)\ket{H}-\sin(\beta)\ket{V}\Big)\ket{L},
\end{equation}
this gives the form
\begin{equation}\label{eq:tan}
\begin{split}
      \sigma_x^w(\beta) &= \frac{\bra{L}\Big(\cos(\beta)\cos^{2N}(\pi/2N)\bra{H}-\sin(\beta)\bra{V}\Big)\Big(\ketbra{D}{D}-\ketbra{A}{A}\Big)\ket{H}\ket{L}}{\bra{L}\Big(\cos(\beta)\cos^{2N}(\pi/2N)\bra{H}-\sin(\beta)\bra{V}\Big)\ket{H}\ket{L}}\\
      &= \frac{-\sin(\beta)}{\cos(\beta)\cos^{2N}(\pi/2N)}\approx -\tan(\beta),
\end{split}
\end{equation}
which goes to 0 as $\beta\rightarrow0$ (or rather, as we postselect on $\ket{H}$), and becomes infinite as $\beta\rightarrow\pi/2$ (as we postselect on $\ket{V})$.

Compare this to the weak value of $\sigma_x$ when the polarising mirror is removed - in this case $\ket{\psi_f}$ remains $\ket{D}\ket{L}$, and so both the numerator and denominator of $\sigma_x$ are $1/\sqrt{2}$, and so $\sigma_x^w=1$. In the extended case, we similarly get $\sigma_x^w(\beta) = \tan(\beta)$ - therefore, the two cases become indistinguishable as $\beta\rightarrow0$ (or rather, as we postselect on $\ket{H}$), and becomes infinitely distinguishable as $\beta\rightarrow\pi/2$ (as we postselect on $\ket{V})$.

Note though that to calculate these weak values in this way, we must already know whether or not the polarising mirror is present in the apparatus.

\section{Experimentally obtaining this weak value}\label{ExpObtWeakVal}

\begin{figure}
    \centering
    \includegraphics[width=0.75\linewidth]{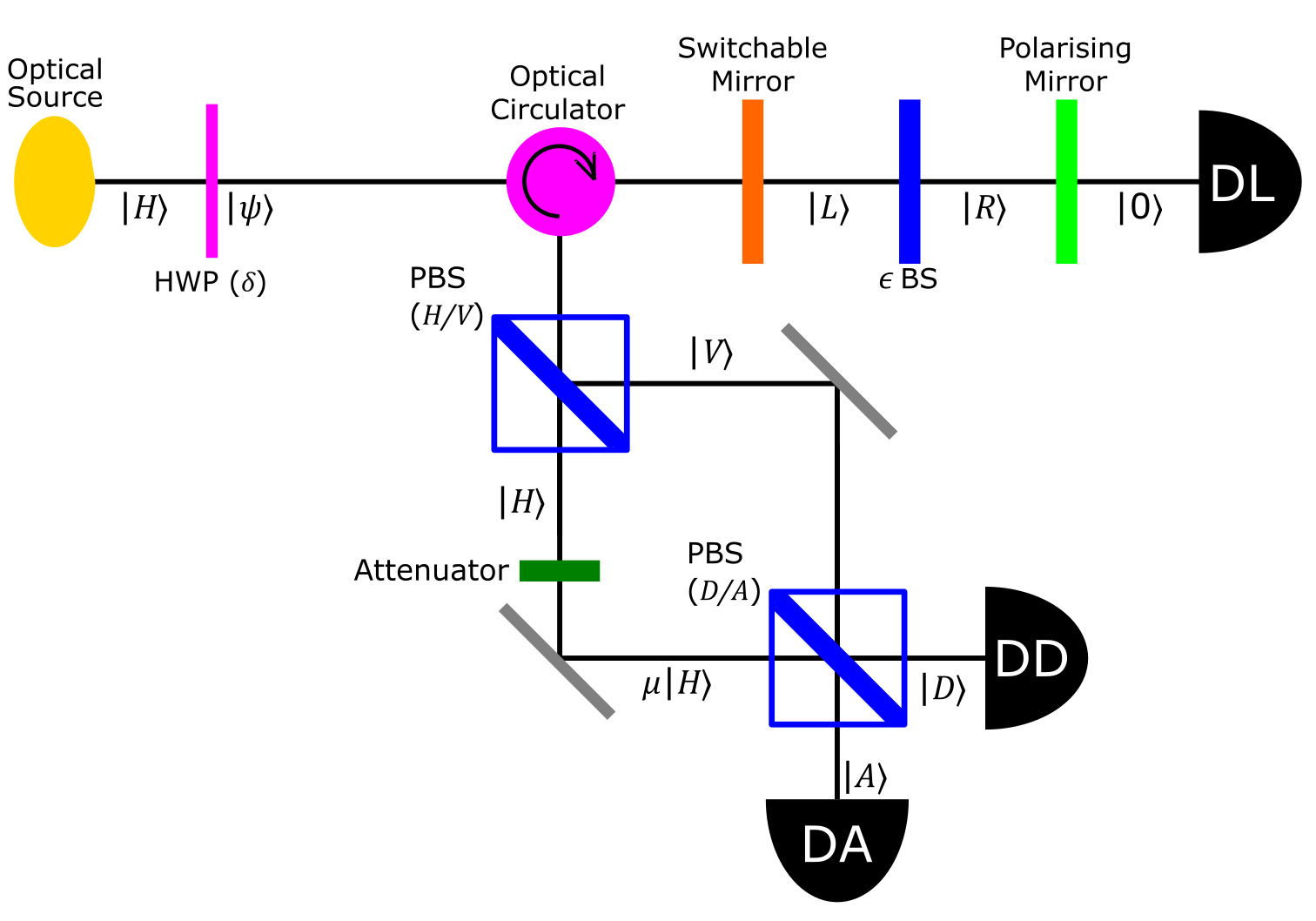}
    \caption{An illustrative alteration to Aharonov et al's protocol to illustrate the relation between the perturbation $\delta$ of our state's polarisation from $\ket{H}$, and the weak value we observe by comparing intensities observed at the detectors. {In this alteration, before the protocol we perturb the polarisation from $\ket{H}$ towards $\ket{V}$ by tiny angle $\delta$, giving us the state $\ket{\delta} = \cos{\delta}\ket{H} + \sin{\delta}\ket{V}$. We then insert this state into our optical analogue of Aharonov et al's device, letting it evolve for $2N$ cycles, before allowing it to leave via the switchable mirror, and travel into the new lower arm via the optical circulator. In this arm, the $\ket{H}$ and $\ket{V}$ components are split by a $H/V$ polarising beamsplitter, before the $\ket{H}$ component is attenuated by attenuation coefficient $\mu$ (to $\mu\cos^{2N}(\pi/2N)\cos(\delta)\ket{H}$). These are recombined at a $D/A$ polarising beamsplitter, which splits off the $D$ and $A$ components, sending these to $DD$ and $DA$ respectively.}}
    \label{fig:John}
\end{figure}

{We now} give an illustrative example to explain why we see a negative weak value when we perform a weak measurement, but cannot see a phase appearing when measured normally. To do this, we describe an experimental proposal to measure the effect indirectly, akin to weak measurement. (We {first} use a direct perturbation in this proposal for clarity and simplicity, {but for those who are interested} give a full treatment for measuring experimentally using weak measurement in Section~\ref{Sect:FullWeak}, and show that this is equivalent to the experimental protocol we give here). We then show this only allows us to observe the effect when we perturb the initial state away from $H$-polarisation, meaning we never observe the effect experimentally in the exact scenario Aharonov et al want us to consider.

In this proposal, we use an altered version of Aharonov et al's apparatus, which we give in Fig.~\ref{fig:John}, where before the protocol we perturb the polarisation from $\ket{H}$ towards $\ket{V}$ by tiny angle $\delta$, giving us the state $\ket{\delta} = \cos{\delta}\ket{H} + \sin{\delta}\ket{V}$. We then insert this state into our optical analogue of Aharonov et al's device, letting it evolve for $2N$ cycles, to state 
\begin{equation}
    \hat{U}^{2N}\ket{\delta} = \cos^{2N}(\frac{\pi}{2N})\cos{\delta}\ket{H} - \sin{\delta}\ket{V},
\end{equation}
before allowing it to leave via the switchable mirror, and travel into the new lower arm via the optical circulator. 

In this arm, the $\ket{H}$ and $\ket{V}$ components are split by a $H/V$ polarising beamsplitter, before the $\ket{H}$ component is attenuated by attenuation coefficient $\mu$ (to $\mu\cos^{2N}(\pi/2N)\cos(\delta)\ket{H}$). These are recombined at a $D/A$ polarising beamsplitter, which splits off the $D$ and $A$ components, sending these to $DD$ and $DA$ respectively. {(This change to a mutually-unbiased basis is necessary to observe the phase difference accrued between the $H$ and $V$-components).}

This means our output state is
\begin{equation}
\begin{split}
       \ket{\delta_f}&= \frac{\mu\cos^{2N}(\frac{\pi}{2N})\cos(\delta)-\sin(\delta)}{\sqrt{2}}\ket{D} + \frac{\mu\cos^{2N}(\frac{\pi}{2N})\cos(\delta)+\sin(\delta)}{\sqrt{2}}\ket{A}.
\end{split}
\end{equation}

By comparing our detection rates $P(D) = \braket{D}{\delta_f}$ and $P(A) = \braket{A}{\delta_f}$, we get
\begin{equation}
    \frac{P(D)-P(A)}{P(D)+P(A)} = \frac{-\sin(\delta)}{\mu\cos^{2N}(\frac{\pi}{2N})\cos(\delta)}.
\end{equation}

When we don't attenuate ($\mu=1$), and set $\delta=\beta$, this is the same as the general form of the weak value, which we give above in Eq. \ref{eq:tan}. Further, by setting our attenuation factor so $\mu=\tan(\delta)$, we can observe the weak value -1 for any value of $\delta$ except $\delta=0$. This is as, when $\delta=0$, our attenuation cause our `reference' $\ket{H}$ value to go to 0, giving us nothing to compare the phase of $\ket{V}$ in our interferometer to and so meaning the relative proportion of intensity detected in $DD$ and $DA$ is balanced ($P(D)=P(A)$). 

What does this have to do with the weak value? While more complex, the entangling of some pointer variable with the $D/A$ polarisation needed to obtain the weak value (see Section~\ref{Sect:FullWeak} to see how this is done experimentally) is equivalent to putting some indeterminate perturbation on the state, taking it from $\ket{H}$ to $\cos{\delta}\ket{H} + \sin{\delta}\ket{V}$ for some indeterminate tiny perturbation $\delta$. By postselecting on some non-zero $\beta$, we force this $\delta$ to be non-zero too, and so observe the effect of Aharonov et al's protocol on some non-zero $\ket{V}$ component (this is shown in more detail in Section~\ref{MethodsProofEntisPert}). This is different from the $\ket{H}$ input state they claim to be evaluating.

While the weak value formula in Eq. \ref{Eq.DQCCWeak} gives a well-defined weak value even without perturbation, experimentally, we cannot see this weak value without perturbing the system - it becomes undefined, as both the difference in detector clicks (the numerator) and the weighting denominator goes to zero. Further, as it is stochastic whether the wavepacket in each case goes to $DA$ or $DD$, and the difference in probabilities between them tends to 0 as $\delta$ goes to 0, our signal-to-noise ratio for experimentally obtaining the weak value goes to 0, and the uncertainty for any finite number of detections becomes far larger than the difference between the weak value reading as -1 or as 1. Therefore, we cannot experimentally tell the difference between the apparatus given in Fig. \ref{fig:DynamCCDiagl}, and a setup only differing through the lack of the spin-dependant right wall, using a wavepacket entirely in state $\ket{\uparrow_z}$. This means we cannot experimentally detect the dynamical quantum Cheshire cat with an unperturbed state.

\section{\texorpdfstring{{Measuring the weak value of $\sigma_x$, and the effect of this on $P_R^w$}}{Measuring the weak value of sigma-x, and the effect of this on PRW}}\label{Sect:FullWeak}

\begin{figure}
    \centering
    \includegraphics[width=0.75\linewidth]{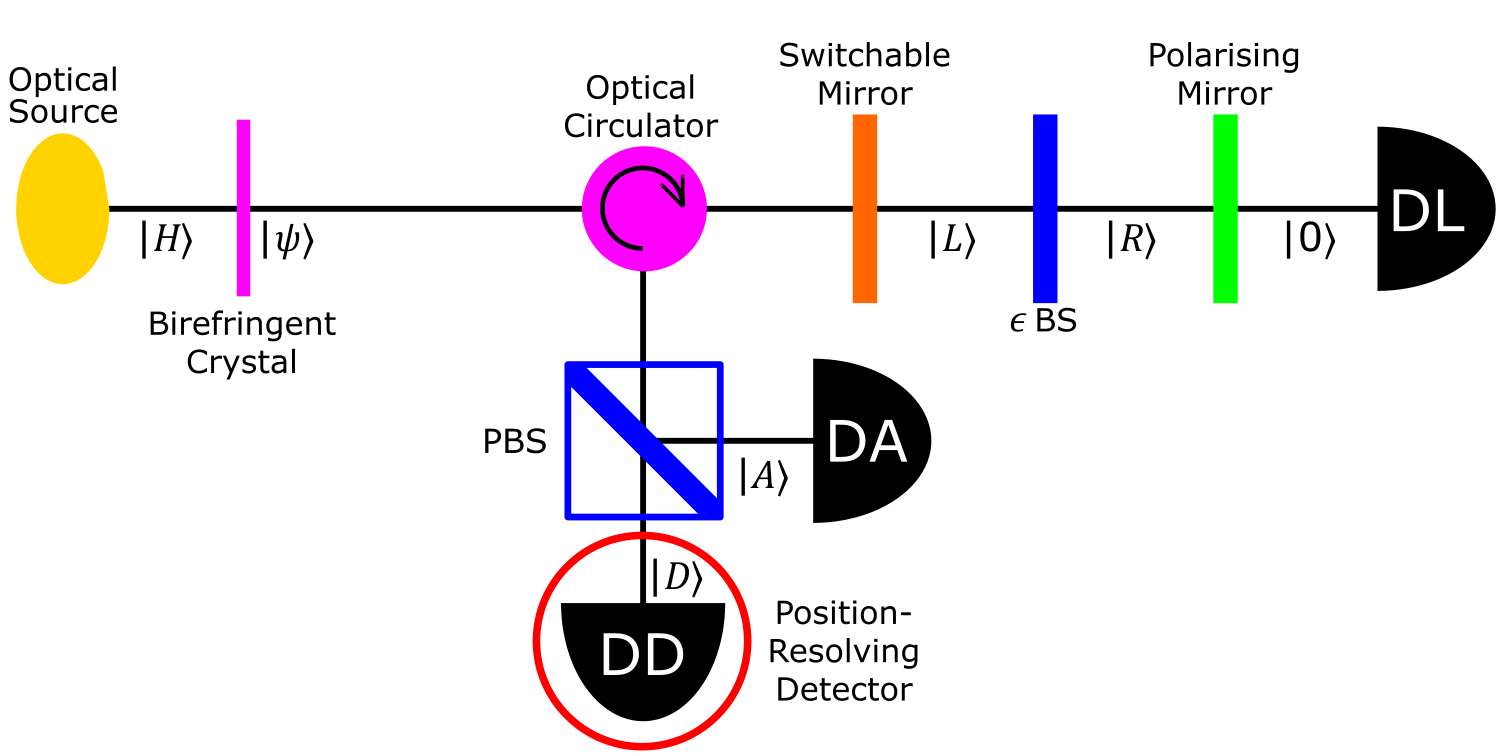}
    \caption{{Adapted form of the apparatus in Fig. \ref{fig:HomodyneDynamCC}, allowing us to obtain the weak value of $\sigma_x$. The red loop indicates the postselection of the wavepacket which goes through the protocol ending up $D$-polarised, and so going to detector $DD$.}}
    \label{fig:WeakDCC}
\end{figure}

{For completeness, we now give an experimental proposal for obtaining the weak value Aharonov et al give for the $\sigma_x$ operator, using a variation of Ritchie et al's protocol for obtaining weak values for polarisation by weakly coupling to the beam centroid position as a pointer state (see Fig.\mbox{\ref{fig:WeakDCC}}) \mbox{\cite{Ritchie1991Realization}}. We then show that the perturbation this causes makes the weak value of the projection operator in the right-hand cavity $\ket{R}$ non-zero for certain $y$-values, illustrating the effect of the weak-measurement induced perturbation. Finally, we use an $H$-polariser in this set-up to show that the entangling between polarisation and position caused by the weak measurement has the same effect as a perturbation of the polarisation - and so we cannot see the dynamical quantum Cheshire cat experimentally if we ensure we only insert $H$-polarised light into the device. Note though that while described physically for the position of the wavepacket centroid, these calculations, and so these conclusions, apply to any case where we use weak measurement to entangle a state with any initially-Gaussian continuous pointer, and so to any weak-value obtainment protocol for the dynamical quantum Cheshire cat. This reinforces the conclusion we come to in Section \mbox{\ref{ExpObtWeakVal}}.}

\subsection{{Obtaining a weak value for \texorpdfstring{$\sigma_x$}{sigma-x} experimentally}}

{To obtain a weak value for $\sigma_x$, we start by describing our wavepacket as a Gaussian in space, with field}

\begin{equation}
    {\ket{E_i} = \sqrt{\frac{2\sqrt{2}}{w_0\sqrt{\pi}}}e^{-\frac{x^2+y^2}{w_0^2}}\Big(\cos(\alpha)\ket{D}+\sin(\alpha)\ket{A}\Big) = \sqrt{\frac{\sqrt{2}}{w_0\sqrt{\pi}}}e^{-\frac{x^2+y^2}{w_0^2}}\Big(\ket{D}+\ket{A}\Big)}
\end{equation}
{(where the second equality uses our preselected input state $\ket{H}$, and so $\alpha=\pi/4$).}

{We then apply a weak measurement on polarisation, using a birefringent-crystalline quartz plate to separate the polarisation-components by a distance $a$ much smaller than the beam width $w_0$, to get}

\begin{equation}\label{Eq:StateEntangledPointer}
\begin{split}
    \ket{E_w} &= \sqrt{\frac{2\sqrt{2}}{w_0\sqrt{\pi}}}e^{-\frac{x^2}{w_0^2}}\Big(\cos(\alpha)e^{-\frac{(y-a/2)^2}{w_0^2}}\ket{D}+\sin(\alpha)e^{-\frac{(y+a/2)^2}{w_0^2}}\ket{A}\Big)/\\
    &= \sqrt{\frac{\sqrt{2}}{w_0\sqrt{\pi}}}e^{-\frac{x^2}{w_0^2}}\Big(e^{-\frac{(y-a/2)^2}{w_0^2}}\ket{D}+e^{-\frac{(y+a/2)^2}{w_0^2}}\ket{A}\Big).
\end{split}
\end{equation}

{If we weren't to apply the protocol, but were instead to send the beam straight to the detector $DA$, this would give the intensity}
\begin{equation}
    {I_w(y) = |\braket{D}{E_w}|^2 = \frac{\sqrt{2}}{w_0\sqrt{\pi}}e^{-\frac{2x^2}{w_0^2}}\Big(e^{-\frac{2(y-a/2)^2}{w_0^2}}\Big),}
\end{equation}
{and so the average beam position at $x=0$}
\begin{equation}
  {\langle y_w\rangle = \int^\infty_{-\infty}y I_w(y) dy = a/2.}
\end{equation}

{If instead we apply the system evolution $\hat{U}^{2N}$ to $\ket{E_w}$, we get the state}

\begin{equation}
\begin{split}
      \ket{E_f}=\hat{U}^{2N}\ket{E_w} =
    &\sqrt{\frac{2\sqrt{2}}{w_0\sqrt{\pi}}}e^{-\frac{x^2}{w_0^2}}\Big(\cos(\alpha)e^{-\frac{(y-a/2)^2}{w_0^2}}\frac{\cos^{2N}(\pi/2N)\ket{H}-\ket{V}}{\sqrt{2}}+\sin(\alpha)e^{-\frac{(y+a/2)^2}{w_0^2}}\frac{\cos^{2N}(\pi/2N)\ket{H}+\ket{V}}{\sqrt{2}}\Big)\\
    \approx&\sqrt{\frac{2\sqrt{2}}{w_0\sqrt{\pi}}}e^{-\frac{x^2}{w_0^2}}\Big(\cos(\alpha)e^{-\frac{(y-a/2)^2}{w_0^2}}\ket{A}+\sin(\alpha)e^{-\frac{(y+a/2)^2}{w_0^2}}\ket{D}\Big)\\
    =&\sqrt{\frac{\sqrt{2}}{w_0\sqrt{\pi}}}e^{-\frac{x^2}{w_0^2}}\Big(e^{-\frac{(y-a/2)^2}{w_0^2}}\ket{A}+e^{-\frac{(y+a/2)^2}{w_0^2}}\ket{D}\Big).
\end{split}
\end{equation}

{Postselecting on the polarisation state $\bra{\beta}=\cos(\beta)\bra{D}+\sin(\beta)\bra{A}$ gives us the amplitude}
\begin{equation}
\begin{split}
     \braket{\beta}{E_f} &= \sqrt{\frac{2\sqrt{2}}{w_0\sqrt{\pi}}}e^{-\frac{x^2}{w_0^2}}\Big(\cos(\alpha)\sin(\beta)e^{-\frac{(y-a/2)^2}{w_0^2}}+\sin(\alpha)\cos(\beta)e^{-\frac{(y+a/2)^2}{w_0^2}}\Big)\\
     &=\sqrt{\frac{\sqrt{2}}{w_0\sqrt{\pi}}}e^{-\frac{x^2}{w_0^2}}\Big(\sin(\beta)e^{-\frac{(y-a/2)^2}{w_0^2}}+\cos(\beta)e^{-\frac{(y+a/2)^2}{w_0^2}}\Big),
\end{split}
\end{equation}
{and so the normalised intensity (at $x=0$) as a function of $y$}
\begin{equation}
\begin{split}
      &I_f(y) =|\braket{\beta}{E_f}|^2_{x=0} =\\
      &\; \frac{2\sqrt{2}}{w_0\sqrt{\pi}} \Big(\cos^2(\alpha)\sin^2(\beta)e^{-\frac{2(y-a/2)^2}{w_0^2}} + \sin^2(\alpha)\cos^2(\beta)e^{-\frac{2(y+a/2)^2}{w_0^2}} + 2\cos(\alpha)\cos(\beta)\sin(\alpha)\sin(\beta)e^{-\frac{2y^2+a^2/2}{w_0^2}}\Big)\\
      &\;= \frac{\sqrt{2}}{w_0\sqrt{\pi}} \Big(\sin^2(\beta)e^{-\frac{2(y-a/2)^2}{w_0^2}} + \cos^2(\beta)e^{-\frac{2(y+a/2)^2}{w_0^2}} + 2\cos(\beta)\sin(\beta)e^{-\frac{2y^2+a^2/2}{w_0^2}}\Big).
\end{split}
\end{equation}

{Treating this as a probability density function, we can get the average position of the centre of the beam on detector $DA$ after the protocol, $\langle y_f\rangle$ as}
\begin{equation}
     {\langle y_f\rangle = \int^\infty_{-\infty}y I_f(y) dy =a(\cos^2(\alpha)\sin^2(\beta)-\sin^2(\alpha)\cos^2(\beta)) = \frac{a}{2}(\sin^2(\beta)-\cos^2(\beta)).}
\end{equation}

{In our protocol, we begin with the state $\ket{H}$ (so $\alpha = \pi/4$), and postselect on state $\ket{D}$ (so $\beta =0$), meaning our average position is $\langle y_f \rangle = -a/2$, and the weak value $\sigma^w_x = \langle y_f\rangle/\langle y_w\rangle = -\frac{a}{2}/\frac{a}{2} = -1$. This is in agreement with the result Aharonov et al give for the weak value of $\sigma_x$. However, more peculiarly, as we change our postselection, we can change the weak value - as $\beta\rightarrow0,\,\pi$ (and so we postselect on either state $\ket{H}$ or $\ket{V}$), this value of $\sigma_x$ goes to 0.}

\subsection{{What if there is no polarising mirror?}}
{If there were no polarising mirror in the apparatus, any light travelling to the right-hand cavity of the apparatus would be lost, regardless of polarisation. Therefore, both $H$ and $V$ components would just get an attenuation of $\cos^{2N}(\pi/2N)$, rather than $V$-polarised components picking up a phase of $-1$. Because of this, $D$-polarised components wouldn't flip to $A$-polarised, but would stay $D$-polarised (and vice-versa), so the beam centroid would stay at $\langle y\rangle=a/2$, and the weak value would be 1, rather than -1. This shows a clear experimentally-detectable difference between when there is and when there isn't a polarising mirror in the right-hand cavity, when we perturb the system to detect this.}

\subsection{{Weak value of \texorpdfstring{$\ketbra{R}{R}$}{the wavepacket being on the right}}}

{Looking at the perturbed state we give above, when we want to find the weak value of the projector $\hat{P}_R = \ketbra{R}{R}$ at some time during the protocol (again, with polarising mirror), we follow Aharonov et al by saying}
\begin{equation}
   {P_R^w = \frac{\bra{L}\bra{D}\hat{U}^{2N-n-1}\hat{U}(T-\tau)\hat{P}_R\hat{U}(\tau)\hat{U}^n\ket{H}\ket{L}}{\bra{L}\bra{D}\hat{U}^{2N}\ket{H}\ket{L}},}
\end{equation}
{but inserting our weakly perturbed state $\ket{E_w(\alpha=\pi/4)}$ instead of the initial $\ket{H}$. For ease, we can say $\hat{U}(\tau) = \hat{\mathds{1}}$, and so $\hat{U}(T-\tau) = \hat{U}$.}

{To see the result we get from this, we must first evaluate the effect of both the $\ket{D}$ and $\ket{A}$ components of this, getting}
\begin{equation}
\begin{split}
    \bra{L}\bra{D}\hat{U}^{2N-n}\hat{P}_R\hat{U}^n\ket{D}\ket{L} =& -\sin(\frac{(2N-n)\pi}{2N})\sin(\frac{n\pi}{2N})/2\\
    \bra{L}\bra{D}\hat{U}^{2N-n}\hat{P}_R\hat{U}^n\ket{A}\ket{L} =& \sin(\frac{(2N-n)\pi}{2N})\sin(\frac{n\pi}{2N})/2,
\end{split}
\end{equation}
{and so}
\begin{equation}
     {\bra{L}\bra{D}\hat{U}^{2N-n}\hat{P}_R\hat{U}^n\ket{E_w}\ket{L} = 
    \sqrt{\frac{\sqrt{2}}{w_0\sqrt{\pi}}}e^{-\frac{x^2}{w_0^2}}\sin(\frac{(2N-n)\pi}{2N})\sin(\frac{n\pi}{2N})\Big(
    e^{-\frac{(y+a/2)^2}{w_0^2}}
     - e^{-\frac{(y-a/2)^2}{w_0^2}}\Big)/2.}
\end{equation}

{From above, we get}
\begin{equation}
   {\bra{L}\bra{D}\hat{U}^{2N}\ket{E_w}\ket{L} = \sqrt{\frac{\sqrt{2}}{w_0\sqrt{\pi}}}e^{-\frac{x^2}{w_0^2}}e^{-\frac{(y+a/2)^2}{w_0^2}}.}
\end{equation}

{Therefore}
\begin{equation}
    {P_R^w = \frac{\bra{L}\bra{D}\hat{U}^{2N-n}\hat{P}_R\hat{U}^n\ket{E_w}\ket{L}}{\bra{L}\bra{D}\hat{U}^{2N}\ket{E_w}\ket{L}}
    =\sin(\frac{(2N-n)\pi}{2N})\sin(\frac{n\pi}{2N})\Big(
    1-e^{\frac{2ay}{w_0^2}}\Big)/2.}
\end{equation}

{While this is 0 at $y=0$, away from this point, it has non-zero weak values. For instance, if we set $y=-a/2$, it reaches its greatest magnitude at $n=N$ of}
\begin{equation}
    {P_R^{w,\text{max}} = \frac{1-e^{-\frac{a^2}{w_0^2}}}{2}.}
\end{equation}

{If we interpret this non-zero weak value of the projection operator $\hat{P}_R$ as showing the presence of some field at a given $y$, this shows the measurable perturbation of the beam centroid by $a$ gives us a non-zero weak value for both polarisation change in the protocol and presence in the right-hand half of the apparatus.}

\subsection{What if we put \texorpdfstring{an $H$-polariser}{a horizontal polariser} between the entangling crystal and the protocol?}\label{MethodsProofEntisPert}
{Elsewhere in the text, we claim that entangling the wavepacket's polarisation with a pointer variable is equivalent to perturbing the polarisation from state $\ket{H}$ by small perturbation $\delta$. In this subsection, we demonstrate this is the case by showing that, when we put a $H$-polariser between the entangling birefringent crystal and the protocol, we cannot observe the weak value experimentally.}

{To do this, we take the state in Eq.~\ref{Eq:StateEntangledPointer}, and apply a $\ketbra{H}{H}$ projector, to get}
\begin{equation}
    {\ketbra{H}{H}\ket{E_w} = \sqrt{\frac{\sqrt{2}}{2w_0\sqrt{\pi}}}e^{-\frac{x^2}{w_0^2}}\Big(e^{-\frac{(y-a/2)^2}{w_0^2}}+e^{-\frac{(y+a/2)^2}{w_0^2}}\Big)\ket{H}.}
\end{equation}

{We then apply the evolution $\hat{U}^{2N}$, which gives}
\begin{equation}\label{Eq.HProjAfter}
\begin{split}
        \hat{U}^{2N}\ketbra{H}{H}\ket{E_w} &= \sqrt{\frac{\sqrt{2}}{2w_0\sqrt{\pi}}}e^{-\frac{x^2}{w_0^2}}\cos^{2N}(\pi/2N)\Big(e^{-\frac{(y-a/2)^2}{w_0^2}}+e^{-\frac{(y+a)^2}{w_0^2}}\Big)\ket{H}+\,\text{Loss}\\
        &\approx \sqrt{\frac{\sqrt{2}}{2w_0\sqrt{\pi}}}e^{-\frac{x^2}{w_0^2}}\Big(e^{-\frac{(y-a/2)^2}{w_0^2}}+e^{-\frac{(y+a)^2}{w_0^2}}\Big)\ket{H}.
\end{split}
\end{equation}

{Finally, we postselect on the wavepacket being $D$-polarised, and so arriving at detector $DD$, to get the intensity function}
\begin{equation}
    {I(y) = \frac{\sqrt{2}}{4w_0\sqrt{\pi}}e^{-\frac{2x^2}{w_0^2}}\Big(e^{-\frac{2(y-a/2)^2}{w_0^2}}+e^{-\frac{(y+a/2)^2}{w_0^2}}+e^{-\frac{2y^2+a^2/2}{w_0^2}}\Big),}
\end{equation}
{and so at $x=0$, the $y$-expectation value is}
\begin{equation}
    {\langle y \rangle = \int^\infty_{-\infty}yI(y)dy = \frac{a}{8}-\frac{a}{8} = 0.}
\end{equation}

{The case we want to compare this to is when there isn't a polarising mirror in the apparatus. In this situation, given this would only apply attenuation of $\cos^{2N}(\pi/2N)$ to the $H$-polarised state, the state after the protocol would be identical to that in Eq. \ref{Eq.HProjAfter}, and so the $y$-expectation value would also be 0. Therefore, there would be no experimentally-detectable difference between the protocol with the polarising mirror, and without the polarising mirror.}

{This proves that, even in cases where it is caused by entanglement between state and pointer (rather than directly as in Section \ref{ExpObtWeakVal}), it is the perturbation of the state away from $H$-polarised which allows us to experimentally detect the difference between the two protocols. While this does not prove that there is no way to perform weak measurement without perturbing the state, we are unaware of any method to do so, and think it unlikely that such a method exists.}

\section{What does this have to do with counterfactual communication?}
Aharonov et al finally claim in their paper that this dynamical quantum Cheshire cat effect is the effect underlying counterfactual communication, an interesting development of exchange-free measurement developed by Salih et al \cite{Salih2013Protocol}. Given the recent interest in counterfactual communication, both in terms of practical applications of quantum computation \cite{Salih2016Qubit,Salih2018Paradox,Salih2020DetTele,Salih2021EFQubit} and metrology \cite{Hance2021CFGI}, and on deeper foundational levels \cite{Salih2018Laws,Aharonov2019Modification,Aharonov2020Nonlocal,Hance2021Quantum}, it is worth examining how our analysis above affects this claim.

The term `counterfactual' in the name `counterfactual communication' comes from Alice (the receiver)'s photon never having travelled to/via Bob (the sender), when she receives a bit of information from Bob in the protocol (i.e. by her postselection on photon either going to her detector $D0$ for a 0-bit, or her detector $D1$ for a 1-bit). Rather than factually interacting with anything on Bob's side, the photon is `counterfactually' telling Alice whether Bob is blocking his channel or not (and so whether his bit-value is 1 or 0 respectively), without ever having been to his channel \cite{Salih2018Laws,Hance2021Quantum}. Therefore, counterfactual communication seems superficially similar to the dynamical quantum Cheshire cat, where there is a difference in the (theoretically-calculated) weak value depending on whether or not a polarisation-dependent mirror is placed at the end of the right-hand cavity, despite the $H$-polarised photon not being able to travel to-and-back-from this mirror.

However, as we have shown above, experimentally determining whether or not the spin-dependent mirror is present requires some small amount of $V$-polarised light to travel to, and back from, the mirror. This $V$-polarised light is obviously not counterfactual. Unlike this, in counterfactual communication, we do not require any small amount of field to travel to-and-back-from Bob's channel to allow Alice to receive information from Bob - anything which goes to Bob is lost (like the $H$-polarised light which travels to the right-hand cavity and is then lost in the unperturbed dynamical quantum Cheshire cat setup).

This means that the dynamical quantum Cheshire cat is not behind Alice being able to detect Bob's blocker setting (and so bit-value) in counterfactual communication. However, counterfactual communication could allow us a way to adapt the dynamical quantum Cheshire cat to make it experimentally detectable without perturbing the initial state.

\section{Conclusion}
We have shown that the dynamical quantum Cheshire cat effect Aharonov et al describe cannot be observed directly via homodyne detection. It can however be detected by obtaining weak values through weak coupling and postselection. This necessarily requires some perturbation away from the initial state, and so requires something to travel through the second cavity, interact with the probed barrier, and return to the first cavity. This shows the protocol is different to counterfactual communication, which does not require any field to travel from Bob (the sender) to Alice (the receiver) for information to be transferred. Further work may try to link this phenomenon instead to contextuality, as has recently been done for the (standard) quantum Cheshire cat protocol \cite{hance2023contextuality}.

\textit{Acknowledgements---}
We thank Sophie Inman, Sandu Popescu and Paul Skrzypczyk for useful discussions. JRH acknowledges support from Hiroshima University's Phoenix Postdoctoral Fellowship for Research, and the University of York's EPSRC DTP grant EP/R513386/1. JRH and JR acknowledge support from the UK Quantum Communications Hub, funded by the EPSRC grants EP/M013472/1 and EP/T001011/1.

\bibliography{ref.bib}
\end{document}